\documentclass[reprint,aps,prl,10pt,letterpaper,superscriptaddress,amsmath,amssymb,floatfix]{revtex4-1}

\usepackage{amsmath,amssymb,bm}
\usepackage{mathtools}
\usepackage{upgreek}
\usepackage{hyperref}
\usepackage{xurl}
\hypersetup{breaklinks=true}

\begin{document}

\title{Observation of the $4f^{14}6s^{2}~^1S_0- 4f^{13}5d6s^{2}(J=2)$ clock transition at 431 nm in $^{171}$Yb}
\author{Akio Kawasaki}
\email{akio.kawasaki@aist.go.jp}
\affiliation{National Metrology Institute of Japan (NMIJ), National Institute of Advanced Industrial Science and Technology (AIST), 1-1-1 Umezono, Tsukuba, Ibaraki 305-8563, Japan}

\author{Takumi Kobayashi}
\affiliation{National Metrology Institute of Japan (NMIJ), National Institute of Advanced Industrial Science and Technology (AIST), 1-1-1 Umezono, Tsukuba, Ibaraki 305-8563, Japan}

\author{Akiko Nishiyama}
\affiliation{National Metrology Institute of Japan (NMIJ), National Institute of Advanced Industrial Science and Technology (AIST), 1-1-1 Umezono, Tsukuba, Ibaraki 305-8563, Japan}

\author{Takehiko Tanabe}
\affiliation{National Metrology Institute of Japan (NMIJ), National Institute of Advanced Industrial Science and Technology (AIST), 1-1-1 Umezono, Tsukuba, Ibaraki 305-8563, Japan}

\author{Masami Yasuda}
\affiliation{National Metrology Institute of Japan (NMIJ), National Institute of Advanced Industrial Science and Technology (AIST), 1-1-1 Umezono, Tsukuba, Ibaraki 305-8563, Japan}


\begin{abstract}
We report on the observation of the $4f^{14}6s^{2}~^1S_0- 4f^{13}5d6s^{2}(J=2)$ transition at 431 nm in $^{171}$Yb by depleting atoms in a magneto-optical trap formed by the $6s^{2}~^1S_0-6s6p~^3P_1$ intercombination transition. The absolute frequency of the transition to the $F=3/2$ state is determined to be $695~171~054~858.1(8.2)$~kHz against physical realization of Coordinated Universal Time maintained by the National Metrology Institute of Japan with a frequency comb. The $g$ factor of the transition to the $F=3/2$ state and the A hyperfine constant are measured to be $g_J=1.54(13)$ and 1123.354(13)~MHz, respectively. More precise spectroscopy of this transition can lead to searches for time variation of the fine structure constant and searches for new physics with isotope shift measurements. 
\end{abstract}

\maketitle


Ytterbium (Yb) is one of the most popular atoms for the purpose of precision spectroscopy. Neutral atoms have two known narrow-linewidth transitions at 507 nm \cite{ApplPhysB.91.57} and 578 nm \cite{PhysRevLett.95.083003}, and Yb$^+$ ions have three narrow-linewidth transitions at 411 nm \cite{PhysRevA.56.2699}, 436 nm \cite{EurophysLett.33.347}, and 467 nm \cite{PhysRevLett.78.1876}. These transitions, as well as some other broader transitions useful for specific applications, are precisely studied based on various motivations, such as operations of optical lattice clocks \cite{Nature.564.87, Metrologia.57.065021, Metrologia.58.055007, Metrologia.57.035007,Metrologia.59.065009,NatPhoton.10.258,Metrologia.57.065017} and ion clocks \cite{PhysRevLett.116.063001,JModOpt.65.585}, searches for the time variation of fundamental constants \cite{PhysRevLett.126.011102}, searches for ultralight dark matter \cite{PhysRevLett.125.201302,PhysRevLett.117.061301,2301.03433,2302.04565,PhysRevLett.129.241301}, diagnosis of quantum degenerate gases \cite{PhysRevLett.101.233002,NewJPhys.18.103009}, searches for atomic parity violation \cite{NatPhys.15.120}, and searches for new physics with isotope shift measurements \cite{PhysRevLett.128.163201, PhysRevLett.125.123002, PhysRevLett.128.073001, PhysRevX.12.021033}. 

In addition to these transitions, three additional narrow-linewidth transitions in neutral Yb atoms are theoretically predicted to be beneficial for fundamental physics \cite{PhysRevA.98.022501,2208.09200,PhysRevLett.120.173001}. The difference between these three transitions is the energy level for the lower-energy state, the $6s^2~^1S_0$ ground state \cite{PhysRevA.98.022501}, $6s6p~^3P_2$ state \cite{2208.09200}, and $6s6p~^3P_0$ state \cite{PhysRevLett.120.173001} that is typically known as the excited state for the clock transition at 578 nm, and they share the same excited state: [Xe]$4f^{13}5d6s^{2}(J=2)$. Because of its unusual electronic configuration where an electron in an inner-shell $4f$ orbital is excited to a $5d$ orbital, transitions connecting this $4f^{13}5d6s^{2}(J=2)$ state and other (meta)stable states have some attractive features. One is high sensitivity to the variation of the fine-structure constant, which is characterized by the sensitivity coefficient of $K=-3.82,~-27,~-15$ for the transition from the $6s^2~^1S_0$ \cite{PhysRevA.98.022501}, $6s6p~^3P_2$ \cite{2208.09200}, and $6s6p~^3P_0$ \cite{PhysRevLett.120.173001} states to the $4f^{13}5d6s^{2}(J=2)$ state, respectively. This also increases the sensitivity to ultralight dark matter through its coupling to the fine structure constant \cite{PhysRevLett.129.241301,PhysRevD.91.015015,2301.03433,2302.04565}. Isotope shift measurements for this transition contribute to new physics searches. Because Yb has a large number of stable bosonic isotopes and various narrow-linewidth transitions in both neutral atoms and Yb$^{+}$ ions, it is one of the best studied atoms to search for new interaction between neutrons and electrons through isotope shift measurements \cite{PhysRevLett.128.163201,PhysRevLett.125.123002,PhysRevLett.128.073001,PhysRevX.12.021033}. One of the important factors for these searches is to utilize transitions involving different electronic structures to vary the average distance between the nucleus and electrons. The configuration of the $4f^{13}5d6s^{2}(J=2)$ state differs from all other states in the transitions investigated so far. Also, high sensitivity to local Lorentz invariance is expected \cite{PhysRevA.98.022501}. However, the transition to the $4f^{13}5d6s^{2}(J=2)$ state from the ground state is not previously reported. 

In this paper, we report an observation of the 431 nm transition from the $6s^{2}~^1S_0$ ground state to the $4f^{13}5d6s^{2}(J=2)$ state in $^{171}$Yb. Among two hyperfine structures, the $F=3/2$ level is carefully characterized. The absolute frequency and magnetic properties of the transition are determined. 

\begin{figure}[!b]
    \includegraphics[width=0.9\columnwidth]{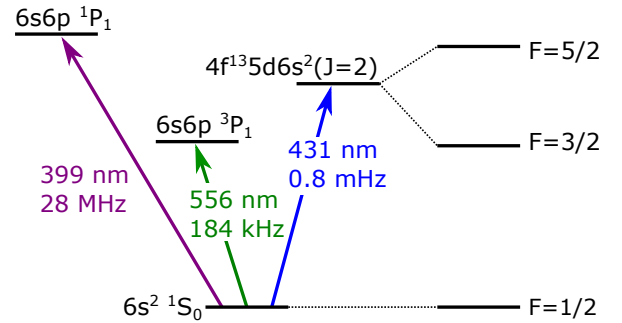}
    \caption{Energy level in a neutral $^{171}$Yb atom related to the search for the $6s^{2}~^1S_0-4f^{13}5d6s^{2}(J=2)$ transition. The right most column shows hyperfine states for the ground state and the $4f^{13}5d6s^{2}(J=2)$ state.}
    \label{PhaseLoss}
\end{figure}

The experimental apparatus was previously used for an optical lattice clock \cite{APEX.2.072501,APEX.5.102401}. Yb atoms emitted from an oven of $\sim 880$ K are slowed down by a Zeeman slower and then trapped in a magneto-optical trap (MOT) generated by the $6s^2~^1S_0-6s6p~^1P_1$ transition at 399 nm (See Fig. \ref{PhaseLoss} for relevant energy levels). Atoms are next transferred to the second-stage MOT formed with the $6s^2~^1S_0-6s6p~^3P_1$ intercombination transition at 556 nm, where $\sim 10^5$ $^{171}$Yb atoms are trapped. After a cooling period, atoms are cooled down to 30 $\upmu$K and maintain the temperature through the period of interrogation by the 431 nm light. 

The 431 nm light, which subsequently illuminates atoms in the MOT, is generated by second harmonics generation from a titanium sapphire (Ti:sapph) laser tuned at 862 nm with a waveguide periodically poled lithium niobate (PPLN). The 862 nm light is frequency-locked to a ULE cavity through a fiber-laser-based frequency comb and a 1064 nm laser. The cavity has short-term relative stability of $\sim2\times10^{-15}$, and its linear frequency drift is canceled by a feedforward, resulting in a long-term stability better than $10^{-13}$ for averaging times of $\lesssim 10^4$ s. The frequency of the 1064 nm light is locked to the ULE cavity by the Pound-Drever-Hall lock, and one of the modes of the frequency comb is phase locked to the 1064 nm laser. The 862 nm laser is phase locked to a different mode of the frequency comb. The linewidth of the 431 nm laser with the lock is $\lesssim1$ Hz, similar to the linewidth of the clock laser used in an Yb optical lattice clock \cite{PhysRevLett.129.241301,OptExpress.7.7891}.

Fluorescence light from the MOT is recorded with a silicon photomultiplier (SiPM) and an amplifying circuit. To suppress unwanted fluctuation in the amount of fluorescence by atoms and background scattering, the power of the 556 nm laser is stabilized by a servo circuit during the interrogation period. 

The initial search for the $4f^{14}6s^{2}~^1S_0-4f^{13}5d6s^{2}(J=2)$ transition is performed with the 431 nm light shone onto the second-stage MOT while its frequency is scanned at the rate of 2~MHz/s by a double-pass acousto-optic modulator (AOM). This 1-s long scan is performed iteratively by shifting the scanning range at 1 MHz steps. To cover a scanning range larger than the modulation bandwidth for the double-pass AOM, the 431 nm laser is frequency-locked to different modes of the frequency comb. The maximum attainable power of the 431 nm light at atoms is $\sim 10$ mW with a beam size focused down to $\sim65$ $\upmu$m $e^{-2}$ radius. When the frequency of the laser is resonant to the $4f^{14}6s^{2}~^1S_0-4f^{13}5d6s^{2}(J=2)$ transition, the amount of the fluorescence from the MOT decreases faster than the intrinsic exponential decay of the second-stage MOT. The $F=3/2$ hyperfine level is first observed with this scan. The $F=5/2$ state is observed with substantially slower scanning rate of 200~kHz/s. The stronger transition for the $F=3/2$ state can be explained by the enhancement in the transition rate due to the hyperfine mixing with the states where the electric dipole transition between the ground state is allowed \cite{PhysRevLett.33.676} in addition to the magnetic quadrupole transition. 

\begin{figure}[!t]
    \includegraphics[width=1\columnwidth]{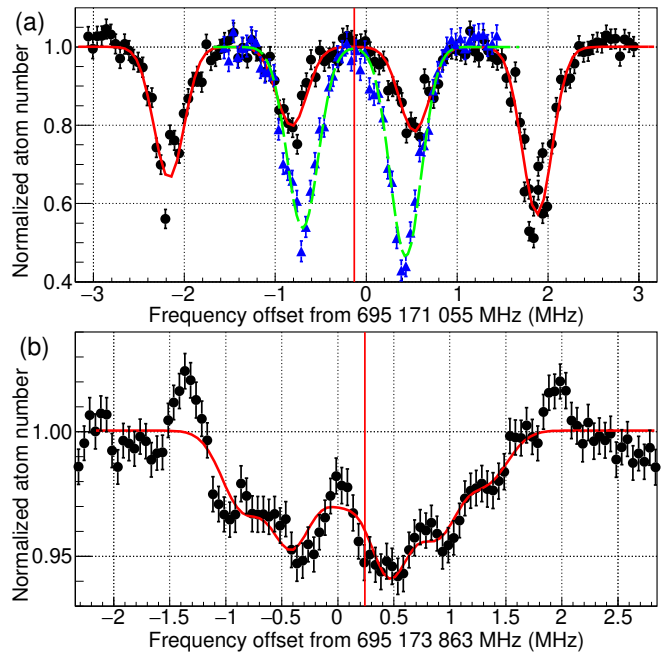}
    \caption{Spectrum of the depletion of the MOT due to the $4f^{14}6s^{2}~^1S_0- 4f^{13}5d6s^{2}(J=2)$ transition: (a) The $F=3/2$ hyperfine state. The black circle (blue triangle) shows the spectrum with 0.62 (0.59) G magnetic field applied in the $y$ ($z$) direction. The red dotted (green dashed) curve is the fit of the black (blue) datapoints. The red vertical line is the average frequency for the four dips obtained by the fit. (b) The $F=5/2$ hyperfine state. The red line shows the fit of the black points. The fitted average frequency of the six dips, shown in the red vertical line, is 0.243(27) MHz. The uncertainty includes the compensation of multiplying square root of $\chi^2/\mathrm{ndf}=1.658$ (ndf: number of degrees of freedom). For all three curves, fits are performed with a constant offset and four (two, six) Gaussians for the $y$ field ($z$ field, $F=5/2$). Gaussians are characterized by their common width, average frequency of four (two, six) dips, spacing between adjacent dips that is regarded as the Zeeman splitting for $\Delta |m_F|=1$, and four (two, six) independent depth of dips. Systematic shifts are compensated.} 
    \label{FigFixedFreqScan}
\end{figure}

Once the initial signature is observed for the $F=3/2$ hyperfine level, spectroscopy while the MOT is turned off for a short period is performed. After the cooling stage of the second-stage MOT, the 556 nm laser is first instantaneously turned off by switching off an AOM, and then the magnetic field is turned off to minimize the extra loss of atoms by accelerating atoms in a certain direction. Atoms are irradiated for 3 ms with the 431 nm light whose frequency is fixed, and then the 556 nm laser and the magnetic field are turned on to recapture the atoms. This cycle of turning the MOT on and off is repeated for 40 times, and the ratio of the atom number after and before these cycles is recorded before rejecting atoms. Each time new atoms are loaded to the MOT, the frequency of the 431 nm light is decremented in a step of 50 kHz to obtain frequency-dependent atom number ratios.

Figure \ref{FigFixedFreqScan} (a) shows an example spectrum. For the $F=3/2$ hyperfine levels, four dips corresponding to four different Zeeman sublevels for the excited state are observed when the bias magnetic field $B_y$ in the $y$ direction is applied, where the $+y$ direction is defined as the direction of the propagation of the 431 nm light and the $z$ axis the vertical axis to which the linear polarization of the 431 nm light is aligned. Note that Zeeman splitting of the ground state in the order of 100 Hz is negligible for the resonance peak broadened to $\sim150$ kHz. 

When the bias magnetic field $B_z$ in the $z$ direction is applied, only $\pi$ transitions are induced, and thus only two peaks are observed. Most of the uncertainty estimates are performed with $B_z$ applied, because the dip is deeper, and $B_y$ is applied in only a few cases. Prior to the measurements, splitting between the four dips is minimized by tuning bias coil currents in the $x$ and $z$ direction to suppress residual magnetic field. When $B_y$ ($B_z$) is applied, obtained spectra are fitted with four (two) Gaussians and a constant offset. 

Even when the four (two) Zeeman sublevels in the $4f^{13}5d6s^{2}(J=2)$ state are not well separated, the fit performs reasonably well, and all data are used to obtain the average frequency of the four (two) resonant frequencies, which is regarded as the resonant frequency for the $F=3/2$ hyperfine level of the $4f^{14}6s^{2}~^1S_0- 4f^{13}5d6s^{2}(J=2)$ transition. After taking an average weighted by the standard deviation of different $B_y$ ($B_z$), the resonant frequency is determined with an uncertainty of a few kilohertz. One major source of this fitting uncertainty presumably comes from the broadened peak of the transition due to the atom's thermal motion. The smallest width of the peak corresponds to the Maxwell-Boltzmann distribution for 41 $\upmu$K. This is slightly higher than the typical temperature of atoms of 30 $\upmu$K measured by absorption imaging, and broader widths are also observed especially for a large bias field, such as in Fig. \ref{FigFixedFreqScan} (a). Potential sources of the broadening are nonuniformity of the residual magnetic field and power broadening. Another source of the uncertainty in this fitting is the shot-to-shot fluctuation in the normalized atom number, as the flat part in Fig. \ref{FigFixedFreqScan}(a) shows. This is separately measured to be at most 0.027, whereas the statistical uncertainty derived within a single shot is on average 0.0085. The error bars in Fig. \ref{FigFixedFreqScan}(a) show overall uncertainties including this shot-to-shot fluctuation. 

Other potential major sources of the uncertainty are the ac Stark shift due to the 431 nm light and the Doppler shift by potential acceleration of atoms when the 556 nm laser is turned off. To test the ac Stark shift, the mean frequency of the two peaks when $B_z$ is applied is measured for different power. The slope obtained from a linear fit gives $-0.75(5.88)$~kHz shift for the 10 mW laser power, which is consistent with zero within the uncertainty. Note that this uncertainty is conservatively assigned to all data, including the measurements performed with lower power.  To evaluate the Doppler shift due to kicking on the MOT, a spectrum taken with a 431~nm probe beam retroreflected is compared with a spectrum without the retroreflection. A shift of 12.4(3.1)~kHz in the single beam case compared to the retroreflected case is observed. With finer scanning with a 1 kHz step, a Doppler free doublet is observed. The average of the mean frequencies of this doublet and the mean frequency of the Doppler broadened peak are 3.8(2.0)~kHz different, and thus conservatively, 3.8~kHz is added to the uncertainty due to the Doppler shift. 

With these major sources of uncertainty, other uncertainties and systematic shifts are negligible. The relative uncertainty of the frequency measurement of the 431 nm laser arising from the measurement noise of the frequency counter around $10^{-13}$ at 1000 s average time, which is on the order of magnitude the same as the overall data acquisition time for the data used for the resonant frequency measurement, is well below 1 kHz. The relative uncertainty of a physical realization of Coordinated Universal Time maintained by the National Metrology Institute of Japan (UTC(NMIJ)) is confirmed to be less than $10^{-14}$ by comparing UTC(NMIJ) with International Atomic Time via a satellite link. Other systematic effects on atoms, such as the second-order Zeeman shift, the black body radiation shift, the dc and ac Stark shift by ambient fields, and the collisional shift typically appear on the level of 1 Hz or less and thus are negligible here. Particularly, it should be noted that the second-order Zeeman shift is predicted to be a few Hz/G$^2$ theoretically \cite{PhysRevA.98.022501}, and this justifies the simple assumption that the Zeeman splitting between adjacent $m_F$ states is the same for all $m_F$ in the $\sim1$ kHz accuracy. The noise added to the laser during the transmission of the 862~nm light through an optical fiber is estimated to be in the order of $10^{-15}$, which is negligible compared to the major sources of uncertainty.

\begin{figure}[!t]
    \includegraphics[width=1\columnwidth]{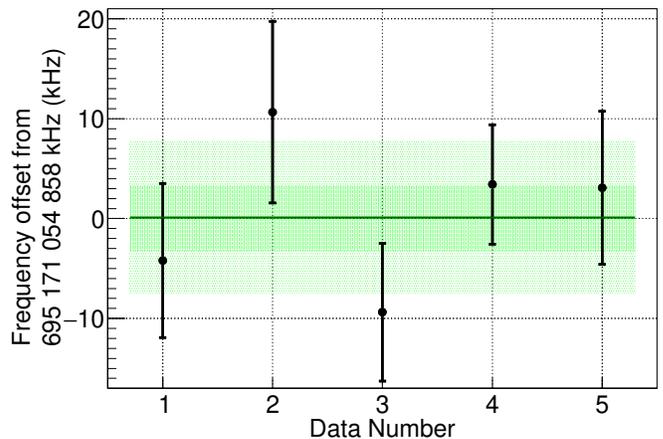}
    \caption{Absolute frequency measurement of the $4f^{14}6s^{2}~^1S_0-4f^{13}5d6s^{2}(J=2)$ transition: green line is the average of five points weighted by the statistical uncertainty. To explain the excessive scattering of the data around average compared to the statistical uncertainty represented by $\chi^2/\mathrm{ndf}=10.3$, the error bar on each data point and thus final statistical uncertainty is inflated by $\sqrt{\chi^2/\mathrm{ndf}}$.  The thick green band is the statistical uncertainty on the average. The light green band is the total uncertainty including the systematic uncertainties. Obtained average in the plot is $0.10\pm3.25(\mathrm{stat})\pm7.65(\mathrm{syst})$ kHz. }
    \label{FigAbsFreq}
\end{figure}

Combining these shifts and uncertainties, the absolute frequency of the $4f^{14}6s^{2}~^1S_0-4f^{13}5d6s^{2}(J=2)$ transition is estimated. The absolute frequency of the 431~nm laser is determined by precisely measuring the repetition frequency and carrier-envelope offset frequency of the frequency comb, which are at 53.148~319~MHz and 30~MHz, respectively, with a frequency counter referenced to UTC(NMIJ). To reduce the measurement noise of the frequency counter, the repetition frequency is measured over $\sim1000$~s, during which different Zeeman shift measurements were performed.  The mode number of the frequency comb is determined by using a wavemeter and an auxiliary frequency comb with a repetition frequency at 50.257~834~5~MHz locked to UTC(NMIJ) \cite{ApplPhysB.73.269}. All other radiofrequency sources for generating frequency offsets between a mode in the frequency comb and the atomic resonance are referenced to UTC(NMIJ). 

Obtained absolute frequencies for several different data set are shown in Fig. \ref{FigAbsFreq}. Obtained average absolute frequency of the transition from the $4f^{14}6s^{2}~^1S_0$ ground state to the $F=3/2$ hyperfine level of the $4f^{13}5d6s^{2}(J=2)$ state is $695~171~054~858.1$~kHz. The fluctuation of the data around the weighted average is substantially larger than their statistical uncertainties ($\chi^2/\mathrm{ndf}=10.3$). The source of this extra uncertainty can be explained as some unknown systematic effect that is specific to the system, e.g., occasional linear drift of the normalized atom number which skews the spectrum. To include all of these into additional statistical uncertainty, $\sqrt{\chi^2/\mathrm{ndf}}$ is multiplied to the uncertainty for the weighted average to obtain overall statistical uncertainty. The same compensation is performed for the ac Stark shift and Doppler shift evaluations. Together with the systematic uncertainty of 7.6~kHz, overall uncertainty is estimated to be 8.2~kHz. This number corresponds to $23~188.410~392~16(27)$~cm$^{-1}$, which has a reasonable agreement with previously reported number of 23188.518~cm$^{-1}$ \cite{SpectroChimicaActaB.35.215,NISTASD}. The uncertainty improved at least four orders of magnitude, and the difference can be explained by an isotope shift. 

\begin{figure}[!t]
    \includegraphics[width=1\columnwidth]{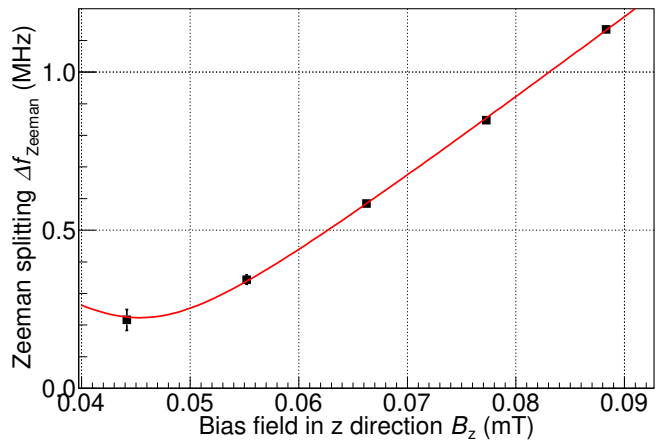}
    \caption{Amount of Zeeman splitting for $\Delta |m_F|=1$ under a different bias magnetic field applied by a bias coil: the red curve is the fit of black points by $\Delta f_{\bf Zeeman}= g_F \mu_B \sqrt{(B_z-B_{z0})^2+B_\perp^2}$, where $g_F$, $B_{z0}$, and $B_\perp=\sqrt{B_x^2+B_y^2}$ are fitted parameters. The best fit provides $g_F=1.85(15)$, $B_{z0}=0.04536(96)$ mT, and $B_\perp=0.00864(97)$ mT. Uncertainties shown here are statistical only, and $\chi^2/\mathrm{ndf}=0.418$ is small enough not to perform any compensation based on this.}
    \label{FigFreqVSZeeman}
\end{figure}

To determine the Land\'e's $g$ factor for the $4f^{14}6s^{2}~^1S_0-4f^{13}5d6s^{2}(J=2)$ transition, a bias magnetic field $B_z$ in the $z$ direction is applied to measure the frequency difference between a pair of $\pi$ transitions between $m_F=\pm1/2$ states. The magnetic field generated by $B_z$ is calibrated by a spectroscopy of the $6s^2~^1S_0-6s6p~^3P_1$ transition, providing an uncertainty of 8 \%. The data shown in Fig. \ref{FigFreqVSZeeman} are fitted with $\Delta f_{\rm Zeeman}= g_F \mu_B \sqrt{(B_z-B_{z0})^2+B_\perp^2}$. The obtained Land\'e's $g$-factor is $g_F=1.85(15)$ and thus $g_J=1.54(13)$, which has a reasonable agreement with a theoretical prediction of 1.5 \cite{PhysRevA.98.022501}.

As for the $F=5/2$ hyperfine state, the transition rate obtained in the setup described in this paper is too small to perform the characterization with iterative 3 ms interrogations with MOT light off. Instead, the 431 nm light is kept on for 1 s without turning off the MOT, during which the frequency of the 431 nm light is scanned over the step size. The ratio of the atom number after and before the 1 s irradiation of the 431 nm light is plotted for different frequencies in Fig. \ref{FigFixedFreqScan}(b). Note that the plot is an average of two scans, with smoothing by averaging five adjacent frequencies. The plot is fitted with a constant offset and six Gaussians, and the mean frequency of the six peaks is regarded as the best estimate for the transition frequency. An additional systematic shift here is the ac Stark shift due to the 556 nm laser. This is separately estimated with the $F=3/2$ hyperfine state to be 0.1018(78) MHz. The absolute frequency of the $F=5/2$ state is therefore determined to be $695~173~863~243(30)$~kHz. The hyperfine splitting for the $^{171}$Yb is estimated to be 2808.385(31) MHz, and thus the A hyperfine constant, which is defined by the magnetic dipole Hamiltonian of $\hat{H}_D=\hat{A} \bm{I} \cdot \bm{J}$ with $\bm{I}$ being nuclear spin and $\bm{J}$ being total electronic angular momentum, is determined to be 1123.354(13)~MHz. 

To summarize, we observed the $4f^{14}6s^{2}~^1S_0-4f^{13}5d6s^{2}(J=2)$ clock transition at 431 nm in $^{171}$Yb, and determined its transition frequency to the $F=3/2$ hyperfine state as $695~171~054~858.1(8.2)$~kHz. This transition has a $g$ factor of $g_J=1.54(13)$, and the hyperfine A constant is measured to be 1123.354(13) MHz. Further investigation of the transition has various significances on fundamental physics, such as new physics searches with isotope shift measurements, dark matter searches with clock comparisons, and searches for time variation of the fundamental constants. For these purposes, more precise spectroscopy is essential. The standard path for this taken in other clock transitions is to trap atoms in an optical lattice for interrogation in Lamb-Dicke regime, and to find a magic wavelength for ac Stark shift free spectroscopy for $<1$ Hz accuracy. Particularly for the new physics search with isotope shift measurements, determination of the isotope shift at this precision is important. 

{\it Note added}--While we are finalizing the manuscript, we noticed another report of the observation of the same transition \cite{2303.09765}, which reports the $g$ factor consistent with our value.

\begin{acknowledgments}
This work was supported by JSPS KAKENHI 21K20359, 22H01161, 22K04942, JST FOREST JPMJFR212S, JST-MIRAI JPMJM118A1, and Research Foundation for Opto-Science and Technology. We are grateful to D. Akamatsu, K. Hosaka, H. Inaba, and S. Okubo for the development of the frequency comb and the stable laser at 1064 nm. 
\end{acknowledgments}

\bibliography{Observing431}

\end{document}